\documentstyle[prd,aps,floats,epsfig]{revtex}

\begin{document}
\draft

\title{\bf Hadronic Decays of $N$ and $\Delta$ Resonances \\ 
in a Chiral Quark Model}
\author{
L. Theussl$^1$, R.F. Wagenbrunn$^2$, B. Desplanques$^1$, W. Plessas$^3$
\vspace{1cm} }
\address{
$^1$Institut des Sciences Nucl\'eaires, \\
(Unit\'e Mixte de Recherche CNRS-IN2P3, UJF),
  F-38026 Grenoble Cedex, France \\
$^2$ Dipartimento di Fisica Nucleare e Teorica, \\
Universit\`a di Pavia and INFN, Pavia 27100, Italy \\
$^3$ Institut f\"ur Theoretische Physik, \\ Universit\"at Graz, 
Universit\"atsplatz  5, A-8010 Graz, Austria 
 }
 
\maketitle

\begin{abstract}
$\pi$ and $\eta$ decay modes of light baryon resonances are investigated
within a
chiral quark model whose hyperfine interaction is based on Goldstone-boson
exchange. For the decay mechanism a modified version of the $^3P_0$  model is
employed. Our primary aim is to provide a further test of the recently proposed
Goldstone-boson-exchange constituent quark model. We compare the predictions for 
$\pi$ and $\eta$ decay widths with experiment and also with results from a
traditional one-gluon-exchange constituent quark model. The differences between
nonrelativistic and semirelativistic versions of the constituent quark models
are outlined. We also discuss the sensitivity of the results on the
parametrization of the meson wave function entering the $^3P_0$  model.
\end{abstract}

\section{Introduction}  

The investigation of hadronic transitions of baryon resonances is currently of
high interest \cite{FBSSUP}. 
On the experimental side, there are considerable efforts
to measure these reactions in order to gain more and improved data on the
resonance states. On the theoretical side, a quantitative description of the
very details of the baryon ground and excited states represents a big challenge
for all hadron models. Obviously the aim is to reach a comprehensive
understanding of the low-energy hadron phenomenology on the basis of
quantum chromodynamics (QCD).

A promising approach to low-energy hadrons consists in constituent quark models
(CQMs). Starting from rudimentary attempts more than two decades ago, one has
constantly improved the description and gained a lot of insight into the
properties of hadrons at low and intermediate energies. Evidently, CQMs can at
most be effective models of QCD in a domain where the fundamental theory is not
(yet) accurately solvable. However, the concept of constituent quarks, in the
beginning mainly motivated by symmetry considerations of hadron multiplets,
nowadays gets more and more justified on the basis of QCD itself
\cite{AOK99,SKUWIL}. It appears that
the spontaneous breaking of chiral symmetry (SB$\chi$S) of QCD is responsible
for the generation of constituent quarks as quasiparticles below a certain
scale. Numerous evidences hint to a chirally broken phase (Nambu-Goldstone mode)
of QCD. 

Recently a chiral constituent quark model (CCQM) has been proposed that exploits
the SB$\chi$S of QCD in deducing the hyperfine interaction of constituent quarks
in light and strange baryons \cite{GLOZ1}. It relies on constituent-quark and
Goldstone-boson fields as the relevant degrees of freedom in an effective
interaction Lagrangian \cite{GLOZ2}. The so-called Goldstone-boson exchange 
(GBE) CQM introduces new symmetry properties into the the hyperfine interaction
of constituent quarks, which are rather different from traditional CQMs
advocating one-gluon-exchange (OGE) dynamics \cite{GLOZ3}. The GBE CQM has turned
out rather successful in producing an accurate description of the whole light
and strange baryon phenomenology in a unified framework \cite{GLOZ1}. 

However, the reproduction of the baryon ground-state and
resonance energies is just one item that has to be fulfilled
by a successful hadron model. In addition, any CQM should
also provide for a description of dynamical properties accessible through all
types of reaction processes. Here we specifically study the performance of the
GBE CQM in hadronic decays of $N$ and $\Delta$ resonances. Thereby we produce a
further test of the reliability of the new kind of hyperfine interaction
based on GBE.

We obtain three-quark wave functions for all the needed ground and excited baryon
states by solving a differential Schr\"odinger-type equation with the stochastic
variational method (SVM) \cite{SUZUK}. These wave functions are then employed within a
modified version of the $^3P_0$  decay model \cite{CANO2} in order to calculate
partial widths for $\pi$ and $\eta$ decays of $N$ and $\Delta$ resonance states
up to $\sim$1.8 GeV. We compare the results to the experimental data and
contrast them to an analogous study along a traditional version of the OGE CQM
\cite{BHAD}. Our main aim is twofold: First we want to see how well the
available data are reproduced by the GBE CQM, and second we wish to find possible
differences between the two distinct types of CQMs.

In the following chapter we give a short description of the quark models used in
the present study. We specify their parametrizations both in a nonrelativistic
and a semirelativistic framework. In chapter 3 we explain the specific decay
model we use here and give the pertinent formulae for the calculation of partial
decay widths. The results are presented in chapter 4 along with a discussion of
their sensitivity on different ingredients both in the CQMs and in the decay
model. Our conclusions are given in chapter 5.

\section{Constituent quark models}
Let us start by specifying the constituent quark models we use in the present
study. The total three-quark Hamiltonian for baryons has the general form
\begin{equation}
\label{hamilton}
H=H_0+V,
\end{equation}
where $H_0$ is the kinetic-energy operator and $V$ contains all the quark-quark
interactions, i.e. confinement plus hyperfine potentials. For constituent
quarks with effective masses of the order of a few hundred MeV the 
kinetic-energy operator should be taken in relativistic form
\begin{equation}
\label{kin_sr}
H^{SR}_0=\sum\limits_{i=1}^3 \sqrt{\vec{p}_i^{\;2}+m_i^2},
\end{equation}
with $m_i$ the masses and $\vec{p}_i$ the 3-momenta of the constituent quarks.
A free Hamiltonian as in Eq. (\ref{kin_sr}) leads to the so-called relativized
or semirelativistic (SR) CQM \cite{CARL}. It helps to avoid pathologies that
usually appear in nonrelativistic constituent quark models \cite{GLOZ3}. We devote
our attention primarily to the SR versions of the GBE and OGE CQMs as described
below. Nevertheless, in order to provide a connection to previous studies of
hadronic baryon decays, we consider also nonrelativistic (NR) versions of the
two types of CQMs, which use the kinetic-energy operator in the form%
\begin{equation}
\label{kin_nr}
H^{NR}_0 = \sum\limits_{i=1}^{3} \left( m_i + \frac{\vec{p}_i\,^2}{2m_i}
\right).
\end{equation}
\subsection{GBE constituent quark model}

For the CCQM relying on GBE dynamics we specifically adhere to 
the version published in Ref. \cite{GLOZ1}. It comes with a mutual quark-quark
interaction 
\begin{equation}
\label{vgbe}
V_{ij}=V_{\rm conf}+V_{\chi},
\end{equation}
with a confinement potential in linear form
\begin{equation}
\label{vconf}
V_{\rm conf}(r_{ij})=V_0+C r_{ij}
\end{equation}
and the chiral interaction consisting of only the spin-spin part of the
pseudoscalar meson exchange
\begin{equation}
\label{vchi}
V_\chi(\vec r_{ij})  =
\left[\sum_{F=1}^3 V_{\pi}(\vec r_{ij}) \lambda_i^F \lambda_j^F\right.
+\sum_{F=4}^7 V_K(\vec r_{ij}) \lambda_i^F \lambda_j^F
\left.\raisebox{0ex}[3ex][3ex]{}+V_{\eta}(\vec r_{ij}) \lambda_i^8 
\lambda_j^8 + \frac{2}{3}V_{\eta'}(\vec r_{ij})\right]
\vec\sigma_i\cdot\vec\sigma_j.
\end{equation}
Here $\vec\sigma_i$ are the Pauli spin matrices and $\lambda_i$
the Gell-Mann flavor matrices of the individual quarks. 
The meson-exchange potentials are parametrized in the form
\begin{equation}
\label{vyuk}
V_\gamma (\vec r_{ij})= \frac{g_\gamma^2}{4\pi}
\frac{1}{12m_im_j}
\left\{\mu_\gamma^2\frac{e^{-\mu_\gamma r_{ij}}}{ r_{ij}}-
\Lambda_\gamma^2\frac{e^{-\Lambda_\gamma r_{ij}}}{ r_{ij}}\right\}\quad
(\gamma=\pi,K,\eta,\eta'),
\end{equation}
with $\mu_\gamma$ the meson masses, $g_\gamma$
the meson-quark coupling constants,
and $\Lambda_\gamma$ the cut-off parameters resulting from the smearing
of the $\delta$-functions (for details see Refs. \cite{GLOZ1} and
also \cite{GLOZ3}). A single coupling constant $g_8$ is taken for
all pseudoscalar octet mesons. In case of the SR GBE CQM it is set 
equal to the pion-quark coupling constant, whose value can be 
deduced from $\pi N$ phenomenology via the Goldberger-Treiman relation.
The coupling constant $g_0$ for the singlet $\eta'$ is
determined differently by a fit to the baryon spectra. The cut-offs
$\Lambda_\gamma$ scale with the phenomenological meson masses 
according to the rule
\begin{equation}
\label{linear}
\Lambda_\gamma=\Lambda_0+\kappa\mu_\gamma.
\end{equation}

The strength and depth of the confinement potential (\ref{vconf})
are determined by $C$ and $V_0$, respectively. While these values have also
been fitted to the baryon spectra, it is interesting to remark that for the
SR GBE CQM the strength $C$ comes out just in consistency with
the QCD string tension. The parameter $V_0$ is needed merely to fix the
ground-state level at the nucleon mass. All the parameter values are
collected in Table \ref{tabgbe}.

\begin{table}[ht]
\caption{\label{tabgbe}
Parameters of the GBE CQM for the semirelativistic \protect \cite{GLOZ1} and
nonrelativistic \protect\cite{FRASC} parametrizations.}
\begin{center}
\begin{tabular}{lcc}
Parameters& SR & NR \\
\hline
\rule{0pt}{3.5ex}
 $\frac{g_8^2}{4\pi}   $      & 0.67 & 1.24  \\
 $(g_0/g_8)^2          $      & 1.34 & 2.23  \\
 $\Lambda_0$ [fm$^{-1}$]      & 2.87 & 5.82  \\
 $\kappa               $      & 0.81 & 1.34  \\
 $C$ [fm$^{-2}$]              & 2.33 & 0.77  \\
 $V_0$ [MeV]                  & -416 &-112  \\
 $m_u=m_d$ [MeV]             & 340 & 340   \\
 $\mu_\pi$  [MeV]             & 139 & 139 \\
 $\mu_\eta$ [MeV]             & 547 & 547 \\
 $\mu_{\eta'}$ [MeV]          & 958 & 958  \\
\end{tabular}
\end{center}
\end{table}

\begin{figure}[ht]
$\begin{array}{lr}&\psfig{file=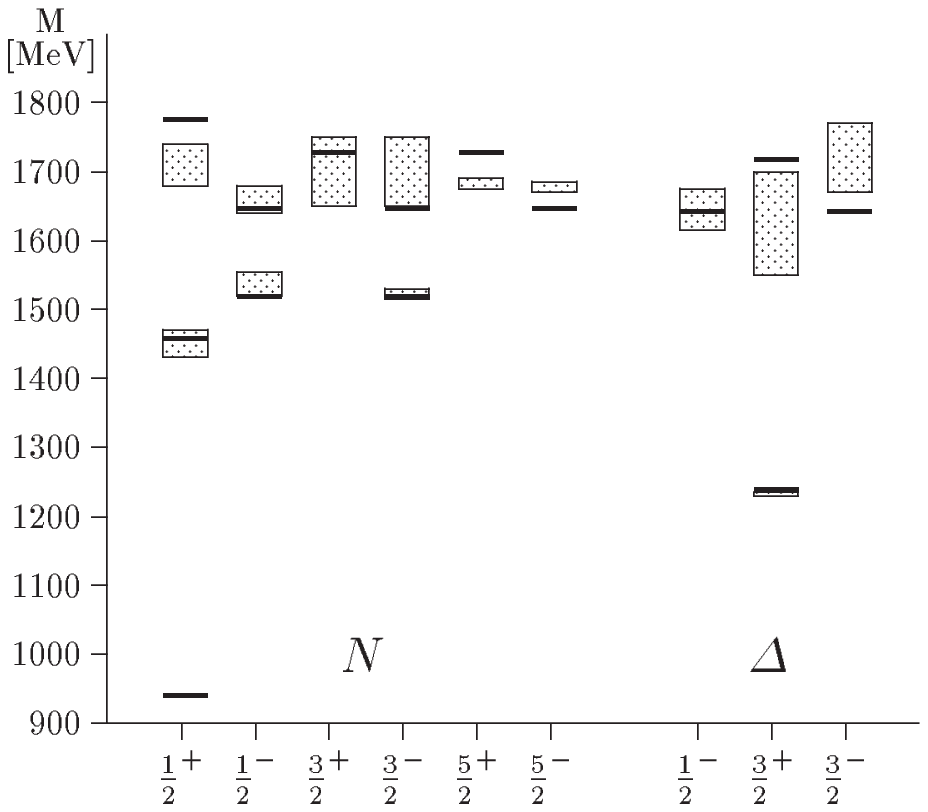,width=7.9cm}
\psfig{file=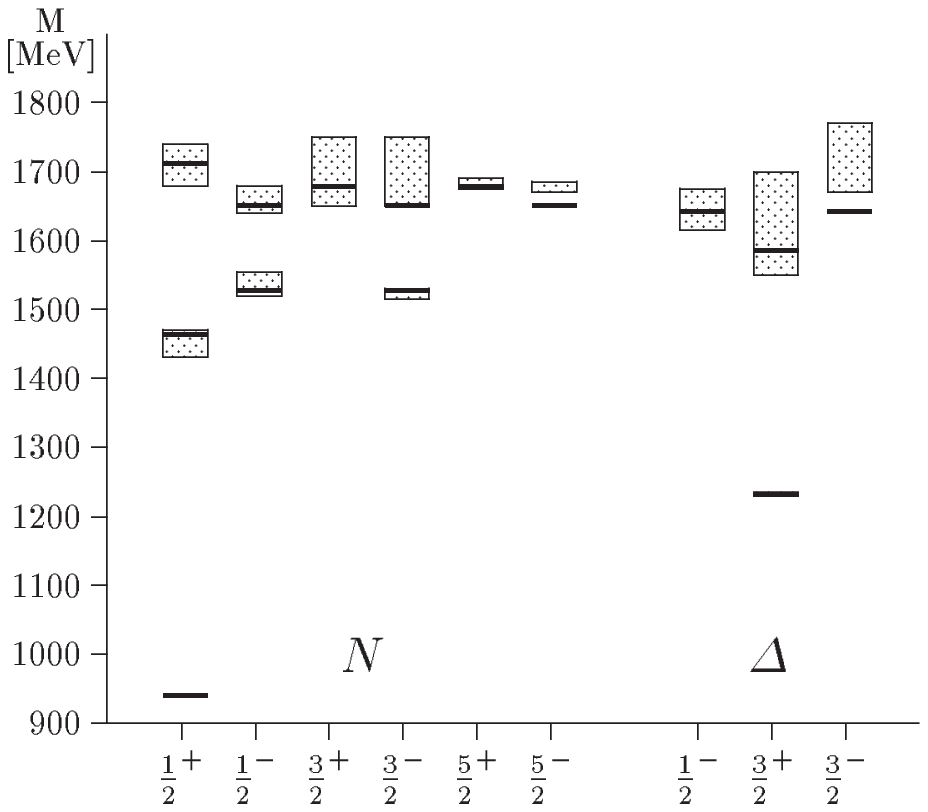,width=7.9cm}\end{array}$
\caption{\label{gbespec}
Energy levels (solid lines) of the lowest $N$ and $\Delta$
states with total angular momentum and parity $J^P$ for
the semirelativistic (left) and the nonrelativistic versions
(right) of the GBE CQM. The shadowed boxes represent the experimental
values with their uncertainties according to the most recent compilation
of the Particle Data Group \protect\cite{PDG}.}
\end{figure}

Table \ref{tabgbe} also contains the parameters for a NR version of the GBE
CQM \cite{FRASC}, i.e. when the potential (\ref{vgbe}) is used
together with the
kinetic-energy operator (\ref{kin_nr}). While the description of the $N$ and
$\Delta$ is achieved with a similar quality 
(cf. Fig.\ \ref{gbespec} and Table \ref{tabmass}), 
it is worthwhile to note the drastically different
values of the fitted parameters (first 6 lines in Table \ref{tabgbe}) in
both the confinement and chiral interactions. In particular, the 
confinement potential becomes unrealistically weak,
while the hyperfine potential gets much enhanced as
compared to the SR case.

\subsection{OGE constituent quark model}

For the purpose of comparison to a different kind of quark-quark
dynamics we employ a traditional OGE CQM. Specifically, it is the model 
following Bhaduri, Cohler, and Nogami (BCN) \cite{BHAD}. 
In this case the total potential has the form

\begin{equation}
\label{bhaduri}
V_{ij}=V_0+C r_{ij}-\frac{2 b}{3 r_{ij}} +\frac{\alpha_s}{9 m_im_j}\Lambda^2
       \frac{e^{-\Lambda r_{ij}}}{r_{ij}}\vec{\sigma}_{i}\cdot\vec{\sigma}_{j},
\end{equation}
i.e. it consists of a short-range Coulomb term, a linear confinement,
and a flavor-independent spin-spin interaction. 
The parameter values for the original BCN potential were determined from a fit
to the meson spectra,  and they were also used in a
previous study \cite{CANO2}.
We have redetermined the model parameters from a
fit to the baryon spectra. Their values
are summarized in Table \ref{taboge}, from where it can be seen
that they differ from the parameter set used in Ref. \cite{BHAD}, 
specifically in the NR case.

\begin{figure}[ht]
$\begin{array}{lr}\psfig{file=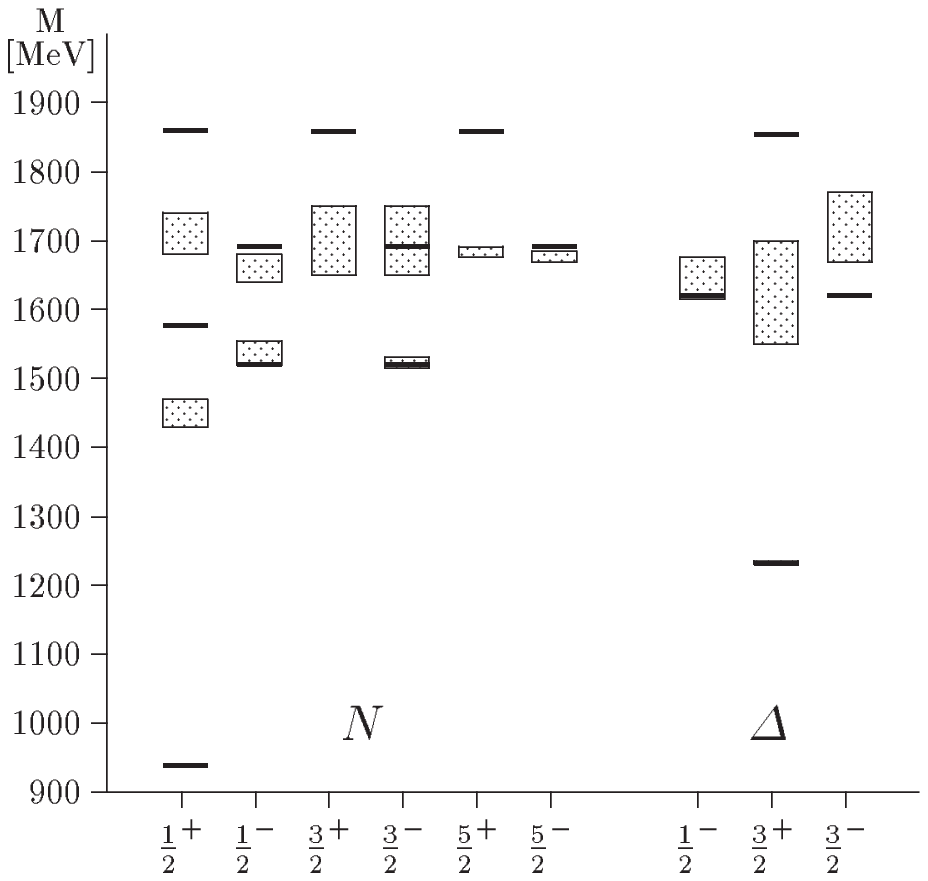,width=7.9cm}&
\psfig{file=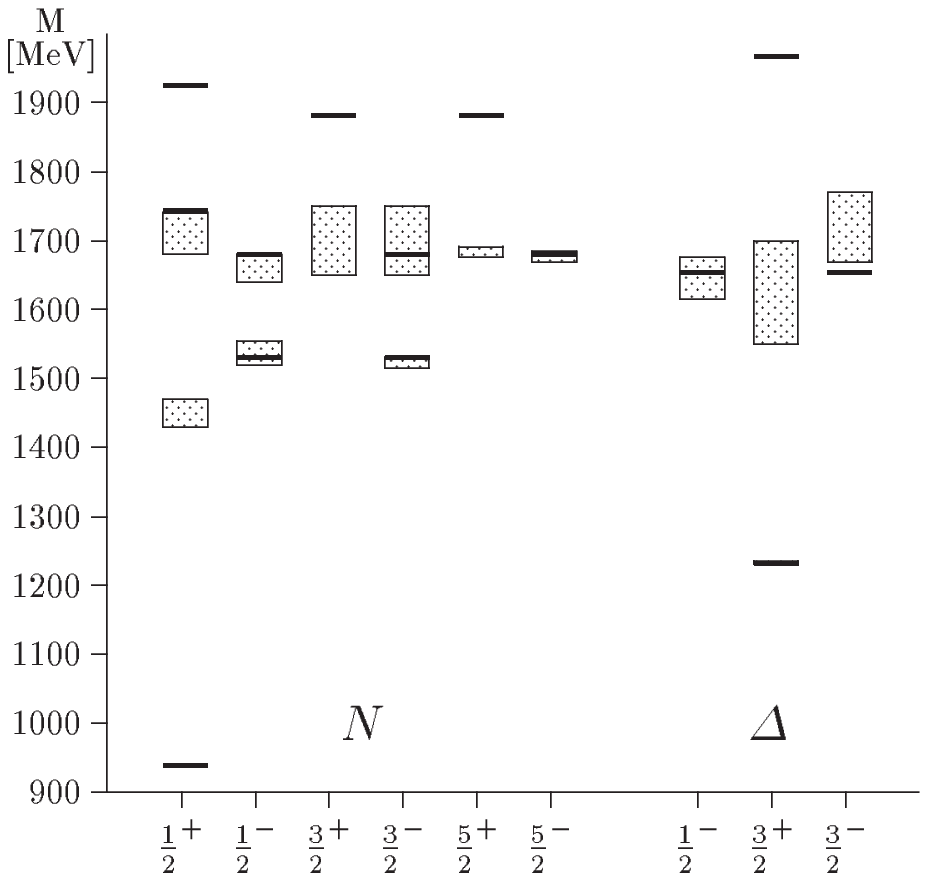,width=7.9cm}\end{array}$
\caption{\label{ogespec}
As Fig.\ \protect\ref{gbespec} but for the OGE CQM after BCN in the
semirelativistic (left) and nonrelativistic versions (right).}
\end{figure}

\begin{table}[ht]
\caption{\label{taboge}
Parameters of the OGE CQM after BCN \protect \cite{BHAD} for the
semirelativistic and nonrelativistic parametrizations.}
\begin{center}
\begin{tabular}{lcc}
Parameters& SR & NR \\
\hline
 $b$                     & 0.57 & 0.825 \\
 $\alpha_s$              & 0.57 & 0.825 \\
 $\Lambda $ [fm$^{-1}$]  & 2.7  & 5     \\
 $C $ [fm$^{-2}$]         & 3.12 & 2.26  \\
 $ V_0 $ [MeV]           & -409 & -366  \\
 $m_u=m_d$ [MeV]         & 337  & 337   \\
\end{tabular}
\end{center}
\end{table}
Again the spectra are produced in quite a similar manner by both the SR and NR
versions (cf. Fig.\ \ref{ogespec} and Table \ref{tabmass}). Of course,
the typical difficulties of OGE CQMs appear, e.g., with respect to the relative
orderings of the lowest positive- and negative-parity excitations.

Before concluding this section a few remarks about the above versions of the
GBE and OGE CQMs are in order. For both cases the models considered here
contain only the most important ingredients for the quark-quark interactions in
baryon spectra, i.e. essentially confinement plus spin-spin hyperfine
interactions. However, both the GBE and OGE models bring about also further
force components, such as central, tensor, and spin-orbit forces.
While their influence must be minor in the $N$ and $\Delta$ spectra (as
demanded by their phenomenological structure) they could be of enhanced
importance in dynamical observables such as hadronic widths, nucleon form
factors, etc. In the study we present below we shall thus essentially explore
the effects of the most prominent parts of the interquark forces. The 
GBE CQM has so far been published only with the spin-spin part of the 
quark-quark interaction \cite{GLOZ1}, \cite{FRASC}. For consistency 
in the comparison, also the OGE CQM is considered only with the 
spin-spin component.

\begin{table}[ht]
\caption{\label{tabmass}
Energies of baryon resonances predicted by the different CQMs 
considered in this work. For all models the nucleon mass is 939 MeV.}

{\begin{tabular}{cccccc}
\rule[-4mm]{0mm}{10mm}$N^*$ & $J^{\pi}$ & 
\multicolumn{4}{c}{M [MeV]}  \\
\hline
\rule[-4mm]{0mm}{10mm}&&  GBE SR & GBE NR & OGE SR & OGE NR  \\
\hline
\rule[-2mm]{0mm}{7mm}$N_{1440} $&$ \frac{1}{2}^+$
                                     & $1459$  & $1465$ 
                                     & $1578$ & $1743$  \\
\rule[-2mm]{0mm}{0mm}$N_{1710} $&$ \frac{1}{2}^+$      
                                     & $1776$    & $1712$  
                                     & $1860$   & $1925$  \\
\rule[-2mm]{0mm}{0mm}$\Delta_{1232}  $&$\frac{3}{2}^+$ 
                                     & $1240$  & $1232$   
                                     & $1232$  & $1232$  \\
\rule[-2mm]{0mm}{0mm}$\Delta_{1600} $&$ \frac{3}{2}^+$ 
                                     & $1718$    & $1585$ 
                                     & $1855$   & $1967$  \\
\rule[-2mm]{0mm}{0mm}$N_{1520} - N_{1535} $&$ \frac{3}{2}^- - \frac{1}{2}^-$     
                                     & $1519$ & $1529$   
                                     & $1521$ & $1531$   \\
\rule[-2mm]{0mm}{0mm}$N_{1650} - N_{1675} - N_{1700} $&$ 
                               \frac{1}{2}^- - \frac{5}{2}^- - \frac{3}{2}^-$      
                                     & $1647$    & $1652$    
                                     & $1691$   & $1681$  \\
\rule[-2mm]{0mm}{0mm}$\Delta_{1620} - \Delta_{1700} $&$ 
                                           \frac{1}{2}^- - \frac{3}{2}^-$ 
                                     & $1642$    & $1642$    
                                     & $1621$    & $1654$   \\
\rule[-2mm]{0mm}{0mm}$N_{1680} - N_{1720} $&$ \frac{5}{2}^+ - \frac{3}{2}^+$     
                                     & $1728$  & $1679$    
                                     & $1858$   & $1883$  \\

\end{tabular}}
\end{table}

\section{The $^3P_0$ model for strong decays}

Investigations of hadronic decays have a long history with first attempts dating
back to the early times of quark models. Still, the definite form of the decay
operator is not yet known. Specific difficulties arising in strong interaction
decays are connected with the extended sizes
of both the baryons and the mesons involved in the decay process. Obviously one
would require a reliable microscopic model that consistently accounts for the
description of both the hadron states and the decay mechanism.

The simplest ansatz for the decay operator is furnished by the elementary
emission model (EEM) \cite{BECCHI,MITRA,FAIMAN}. Therein a pointlike meson is
produced by a single constituent quark in the decaying baryon state. Evidently,
this assumption leads to shortcomings, as found in a number of investigations
with various CQMs (cf., for example, ref. \cite{STAST1}). A preliminary study of
baryon decays for the relativistic GBE CQM along the EEM was performed in ref.
\cite{GLOZ4}.

An improved description of hadron decays is provided by the $^3P_0$  (or quark-pair
creation) model. Here a $q\bar{q}$  pair is created from the vacuum and by a subsequent
rearrangement the final meson and baryon states are produced. The $^3P_0$  model
naturally allows to implement the extended structure of the emitted meson. By
definition, the quark-antiquark pair must carry the quantum numbers of the
vacuum, i.e. it is a color and flavor singlet, has positive P- and C-parity,
total angular momentum $J=0$ and carries total linear momentum zero. From
$P=-(-1)^L$ and $C=(-1)^{L+S}$ one deduces as the simplest choice $L=S=1$. The
corresponding transition operator for the decay can thus be expressed as 
\cite{YAOUAN}

\begin{equation}
\label{3P0op}
{\cal T}  = 
\gamma \sum_{i,j} \int d\vec{p}_q \, d\vec{p}_{\bar{q}} \,
\delta(\vec{p}_q + \vec{p}_{\bar{q}}) 
\sum_{m}\left[C_{1m1-m}^{00}
{\cal Y}_1^m(\vec{p}_{\bar{q}} - \vec{p}_{q})
(\chi_1^{-m}(i,j)\phi_0(i,j)
\right] 
b_{i}^{\dagger}(\vec{p}_{q})\, d_{j}^{\dagger}(\vec{p}_{\bar{q}})
\end{equation}

\noindent
where, in evident notation, the momenta refer to the quark and antiquark states
created by the operators $b_{i}^{\dagger}$ and $d_{j}^{\dagger}$, respectively.
${\cal Y}_{L}^{M}(\vec{p}\,)=
p^L Y_L^M (\hat{p})$ is a solid harmonics function, which gets
coupled with the triplet spin wave function $\chi$ to give $J=0$.
$\phi_0$ is the flavor singlet wave function and the summation $\sum_{i,j}$ runs
over spin and flavor indices.
The pair-creation constant $\gamma$ is a
dimensionless coefficient which is the only adjustable parameter of the model
(apart from factors entering an eventual parametrization of the meson
wave functions). Note that in Eq. (\ref{3P0op}) we have omitted a factor 3 in
front of this constant which is frequently used to cancel a factor $1/3$
coming from the matrix element of color wave functions, which are not 
written out explicitly here.

The transition matrix element for the
process $B \rightarrow B' M$ is then expressed as

\begin{equation}
\label{3p0matrix}
\langle B' M \mid {\cal T} \mid B \rangle \equiv
\langle B' M \mid H \mid B \rangle = 
 3 \gamma \sum_{m}C_{1m1-m}^{00} \, {\cal I}_m =: 
 \delta(\vec{P}-\vec{P}'-\vec{q}) {\cal A}
\end{equation}  

\noindent
Here, the factor $3$ comes from the different possibilities of rearranging the
quarks in the initial and final state, taking into account the symmetry of the
wave functions.
The momentum integral of Eq. (\ref{3p0matrix}) takes the form

\begin{eqnarray}
\label{I_m1}
{\cal I}_m &=& \int d\vec p_1\,d\vec p_2\,d\vec p_3\,
 d\vec p_4\,d\vec p_5 
 {\cal Y}_1^m(\vec{p}_4 - \vec{p}_5)\
 \delta(\vec{p}_4 + \vec{p}_5)\ \Phi_{pair}^{-m} \nonumber \\
  & & \qquad
\left[\Psi_{B'}(\vec p_1,\vec p_2,\vec p_4)\,\Phi_{B'}\right]^{*}
\left[\Psi_{M}(\vec p_3,\vec p_5)\,\Phi_{M}\right]^{*} 
\left[\Psi_{B}(\vec p_1,\vec p_2,\vec p_3)\,\Phi_{B}\right].
\end{eqnarray}
Here, $\vec p_1,\vec p_2,\vec p_3$ are the individual quark momenta of the
initial baryon B which sum up to a total momentum 
$\vec{P}=\sum_{i=1}^{3}\vec{p}_i=0$ in the
rest frame of B. The meson carries away the momentum $\vec{q}=\vec p_3+\vec
p_5$, and the residual baryon B$'$ has momentum 
$\vec{P}'=\vec p_1+\vec p_2+\vec p_4= -\vec{q}$, due to momentum conservation in
the decay process. Finally, we denoted the combined spin-isospin wave functions
involved in the decay process by $\Phi$.

In a next step, one separates the center-of-mass and relative
motions in all hadron wave functions, what permits to carry out some of the
integrations in Eq. (\ref{I_m1}):

\begin{eqnarray}
\label{I_m2}
{\cal I}_m &=& \delta(\vec{P}-\vec{P}'-\vec{q}) 
\int d\vec p_x\,d\vec p_y\,
 {\cal Y}_1^m(2\vec{q} + 2\vec{p}_y)\
 \Phi_{pair}^{-m} \nonumber\\
  & & \qquad
\left[\Psi_{B'}(\vec{p}_x,\frac{2}{3}\vec{q}+\vec{p}_y)\,\Phi_{B'}\right]^{*}
\left[\Psi_{M}(-\frac{1}{2}\vec{q}-\vec{p}_y)\,\Phi_{M}\right]^{*} 
\left[\Psi_{B}(\vec{p}_x,\vec{p}_y)\,\Phi_{B}\right],
\end{eqnarray}
where $\vec{p}_x=\frac{1}{2} (\vec{p}_1 - \vec{p}_2)$ and 
$\vec{p}_y = \frac{1}{3}(2\vec{p}_3-\vec{p}_1 - \vec{p}_2)$ are the momenta
conjugate to the Jacobi coordinates $\vec{x}$ and $\vec{y}$.

In Ref. \cite{CANO2}, it was observed that the $^3P_0$  model can be modified so as
to reproduce the EEM in the limit of a pointlike meson. 
Taking also into account a relativistic boost effect, this requires the
replacements

\begin{eqnarray}
\label{replacement1}
\gamma & \longrightarrow & \gamma \sqrt{\frac{\mu}{\omega}} \\
\label{replacement2}
{\cal Y}_1^m (2\vec{q} + 2\vec{p}_y)
& \longrightarrow & 
{\cal Y}_1^m \left([1+\frac{\omega}{2m}]\vec{q} + 
\frac{\omega}{m}\vec{p}_y\right),
\end{eqnarray}
where $\mu$ is the mass of the emitted meson and 
$\omega=\sqrt{\mu^2+\vec{q}\,^2}$ its energy. 

The partial decay width is then obtained by

\begin{equation}
\label{width}
\Gamma = \frac{1}{\pi}\, 
\frac{q \, E \, \omega}{M_B} 
\mid   {\cal A}  \mid ^{2},
\end{equation}
where $M_B$ is the mass of the decaying resonance, $E$ the energy of the
final state baryon, and ${\cal A}$ is defined by Eq. (\ref{3p0matrix}).
In Eq. (\ref{width}) one still has to sum over final and to average over
initial spin-isospin channels. 

For the meson wave function in configuration space we first adopt 
a simple parametrization of the Gaussian type

\begin{equation}
\label{gauss}
\Psi_G (\vec{r}) = \frac{1}{(\pi R^2)^{3/4}} 
exp \left(-\frac{r^2}{2R^2}\right),
\end{equation}
where the parameter $R^2$ is related to the mean square radius of the meson by 
$\langle r^2 \rangle =\frac{3}{2} R^2$.
While facilitating the calculations, this choice certainly cannot
be regarded as a realistic representation of a meson wave function. We shall
therefore investigate the influence of a different analytic form of meson
wave functions on the baryon decay widths.

From the Fourier transform of the electromagnetic pion form factor, one can
deduce a pion wave function that takes a Yukawa-like form:

\begin{equation}
\label{yukawa}
\Psi_Y (\vec{r}) = 
 \frac{1}{\sqrt{4\pi}}\frac{m}{\sqrt{r}}exp\left( -\frac{m r}{2}\right).
\end{equation}
Here the parameter $m$ is related to the mean square radius of the meson by 
$\langle r^2 \rangle =6/m^2$.
Even if it is not physically meaningful, this expression may serve
as a comparison to the Gaussian form.

\begin{figure}[ht]
\begin{center}
$\begin{array}{lr}
\psfig{file=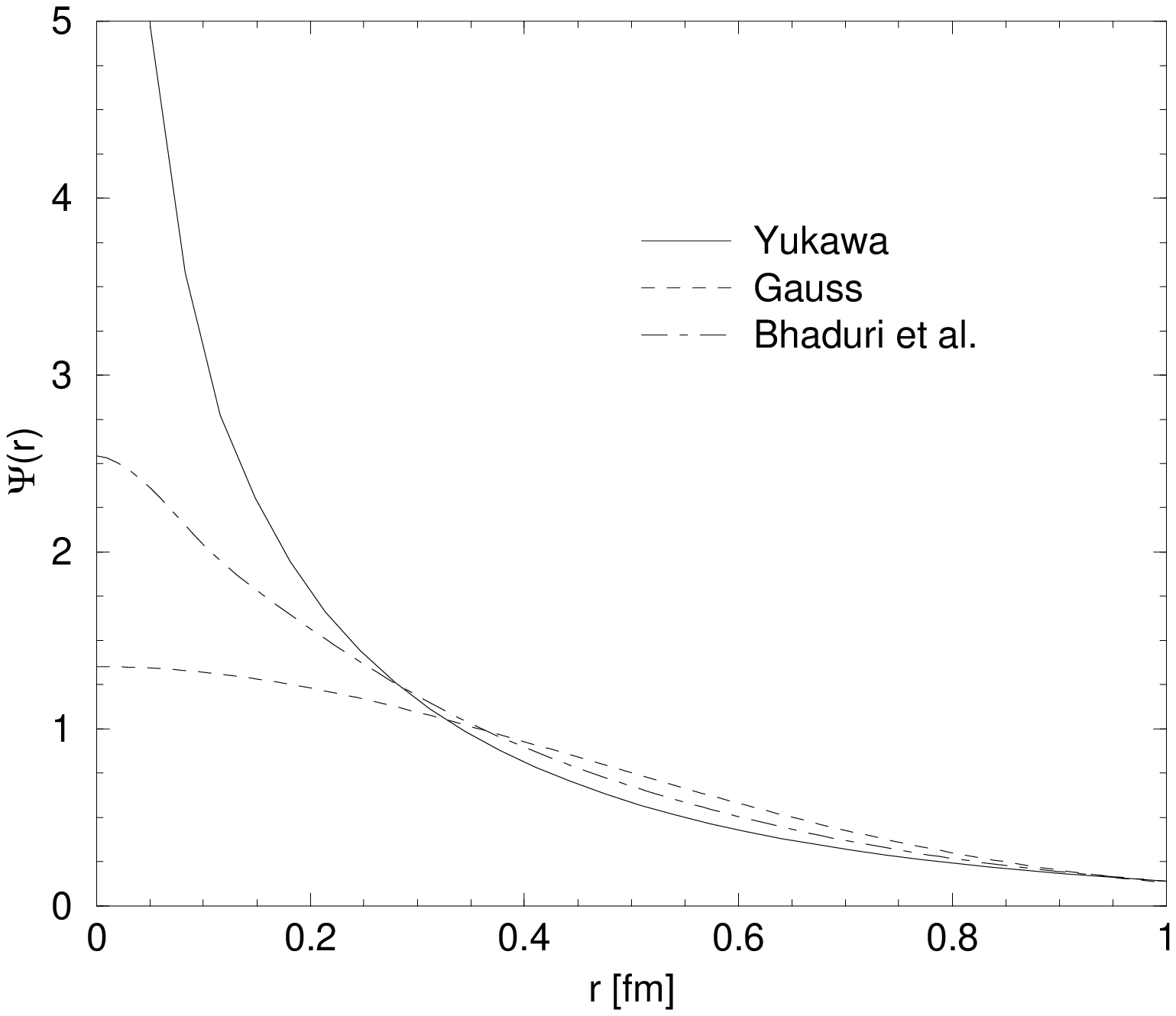,width=16em}&
\psfig{file=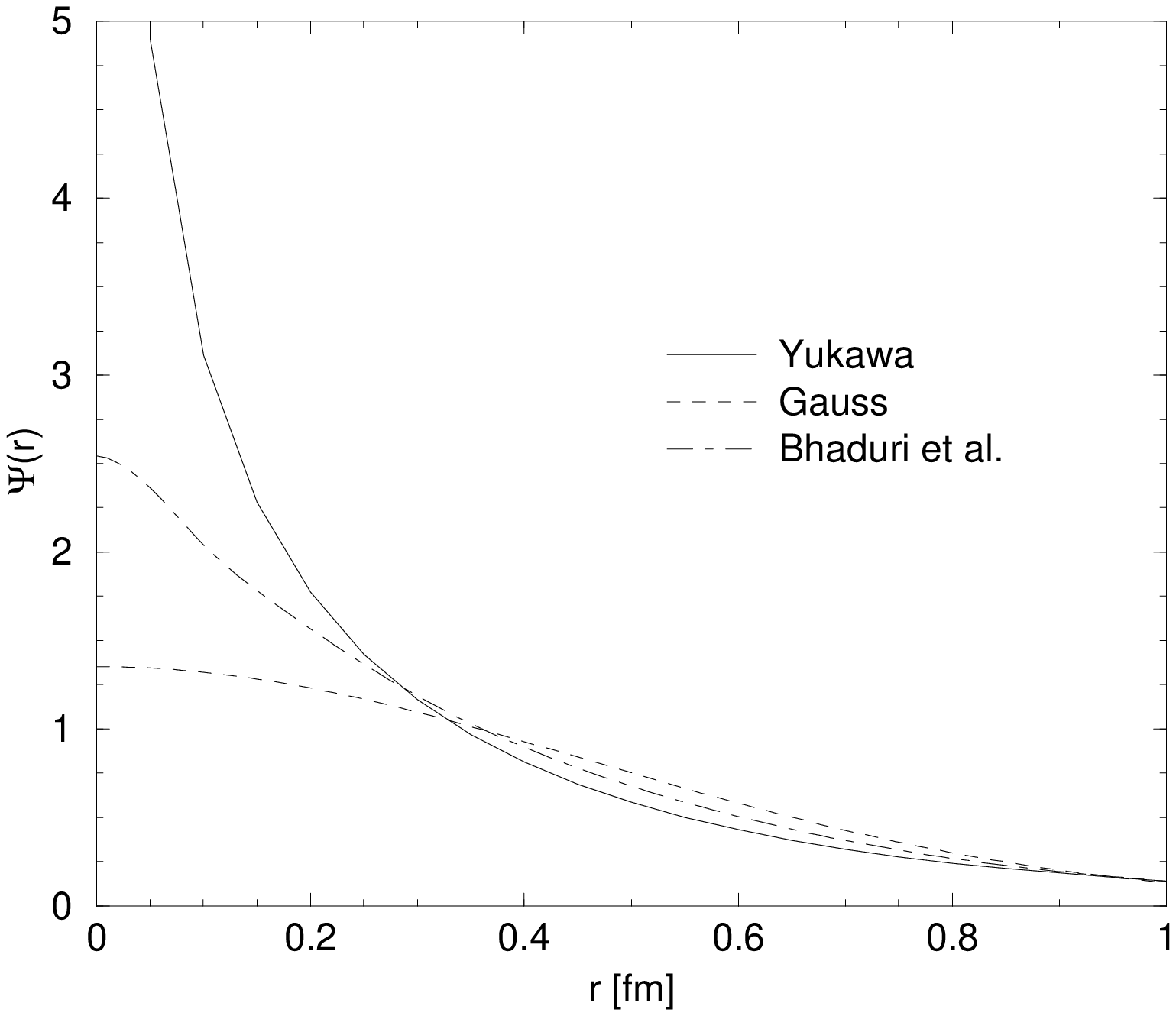,width=16em}
\end{array}$
\caption{\label{ygbwf}
Meson wave functions in momentum (left) and configuration space (right).
Gaussian and Yukawa forms are compared to the exact wave function following from
the original quark-antiquark potential of Bhaduri et al.\protect\cite{BHAD}.}
\end{center}
\end{figure}

A graphical representation of the meson wave functions is given in
Fig.\ \ref{ygbwf},
where we compare the above forms to the wave function that follows from the
original potential of Bhaduri et al. \cite{BHAD}. 
It can be seen that the exact wave function lies just between the extreme 
choices of a Yukawa and a
Gaussian form. The parameters of Eqs. (\ref{gauss}) and (\ref{yukawa}) have
been fitted to give the same root mean square radius for the pion as the
wave function from the potential of ref. \cite{BHAD},
that is $r_{\pi}=0.565$ fm. For simplicity we use the same
parametrization for the wave function of the $\eta$ meson.

\section{Results for $\pi$ and $\eta$ partial decay widths}

In this section we shall present the results for the $\pi$ and $\eta$
decay modes of $N$ and $\Delta$ resonances, as predicted by the CQMs
specified in Sec. 2. At the beginning we discuss some features of the baryon
wave functions.

\subsection{Three-quark wave functions}

The solutions of the three-quark Hamiltonians have been obtained by solving
the corresponding Schr\"odinger-type differential equations with the SVM
\cite{SUZUK}. The accuracy that is attained with respect to the eigenenergies 
in Table \ref{tabmass} is generally within a few percent even for the highest 
states considered. In this context the SVM was carefully conterchecked with 
complementary approaches, such as the Faddeev method \cite{GLOZ3,PKPLE}. 
Another measure for the accuracy of the solution of the three-quark problem is 
the mean square radius of the wave function. In
the context of the present work this quantity is also useful for understanding 
some general characteristics of the results for decay widths, 
which are connected to the baryon sizes. 
In Table \ref{tabrms} we therefore quote mean square radii of the 
$N$ and $\Delta$ ground state wave functions for the CQMs considered. 
The values
refer to the case with pointlike constituent quarks. Therefore they are 
probably not realistic and must not be compared to experimental values. 
They are only useful to get
insight into the relative extensions of the wave functions from each CQM.

\begin{table}[ht]
\caption{\label{tabrms}
Mean square radii of $N$ and $\Delta$ from the various CQMs, 
assuming pointlike constituent quarks.}
\begin{tabular}{ccccc}
\rule[-4mm]{0mm}{10mm} &  GBE SR & GBE NR & OGE SR & OGE NR  \\
\hline
\rule[-5mm]{0mm}{10mm} $\left< r_N^2 \right>$ [fm$^{2}$]&
0.092 & 0.134 & 0.076 & 0.219   \\
\rule[-2mm]{0mm}{0mm} $\left< r_{\Delta}^2 \right>$ [fm$^{2}$] &
0.152 & 0.172 & 0.115 & 0.288   \\
\end{tabular}
\end{table}

Obviously, the values of the mean square radii are all rather small. 
Within each type of CQM, GBE or OGE,
they are smaller in the semirelativistic cases, as it was already
observed in ref. \cite{CARL}. 
This may be viewed as a consequence of the
stronger confinement generally needed in the semirelativistic CQMs. Inspection
of the absolute magnitudes of the relevant quantities in Tables 
\ref{tabgbe}, \ref{taboge}, and \ref{tabrms} shows, however, that
confinement cannot be the only factor determining the mean square radii of
the wave functions (note that the differences in confining strengths 
are much larger for the GBE parametrizations). 
A smaller extension of the baryon wave functions evidently implies
even larger values for internal momenta (in the semirelativistic CQMs).
This will help to explain certain
results for decay widths involving high momenta in the next sections.

\subsection{$\pi$ decays}

The results for the partial widths of the $\pi$ decay modes 
of the $N$ and $\Delta$ resonances are shown in Table \ref{tab1}.
All values have been calculated with the Gaussian-type parametrization of the
meson wave function of Eq. (\ref{gauss}). For the baryons, the theoretical
masses have been used as predicted by the different CQMs in Table \ref{tabmass}. 
In each case the strength parameter $\gamma$ introduced into the decay
operator in Eq. (\ref{3P0op}) has been adjusted so as to reproduce the 
$\Delta_{1232} \rightarrow N\pi$ decay width. All the other decay widths can
then be considered as genuine predictions of the CQMs along the modified
$^3P_0$ model.

\begin{table}[ht]
\caption{\label{tab1}
Decay widths of baryon resonances for the GBE and OGE constituent 
quark models both in nonrelativistic and 
semirelativistic parametrizations. 
A Gaussian-type meson wave function with $r_{\pi}=r_{\eta}=0.565$ fm
was used along with a modified $^3P_0$ decay model.
Experimental data are from ref. \protect\cite{PDG};
for the quoted uncertainties refer to the text.}

\begin{tabular}{cccccccccccc}
\rule[-4mm]{0mm}{10mm}$N^*$ & $J^{\pi}$ & 
\multicolumn{5}{c}{$\Gamma(N^*\rightarrow N\pi$) [MeV]} &
\multicolumn{5}{c}{$\Gamma(N^*\rightarrow N\eta$) [MeV]} \\
\hline
\rule[-4mm]{0mm}{10mm}&&  GBE SR & GBE NR & OGE SR & OGE NR & Exp.
&  GBE SR & GBE NR & OGE SR & OGE NR & Exp. \\
\hline
\rule[-2mm]{0mm}{7mm}$N_{1440} $&$ \frac{1}{2}^+$
                                     & $517$  & $258$ 
                                     & $1064$ & $161$ 
                                     & $(227\pm18)^{+70}_{-59}$ 
				     &  &  & 6 & 10 & \\
\rule[-2mm]{0mm}{0mm}$N_{1710} $&$ \frac{1}{2}^+$      
                                     & $54$    & $14$  
                                     & $202$   & $8$    
                                     & $(15\pm5)^{+30}_{-5}$ 
				     & $26$   & $4$  
                                     & $50$   & $10$ &  \\
\rule[-2mm]{0mm}{0mm}$\Delta_{1232}  $&$\frac{3}{2}^+$ 
                                     & $120$  & $120$   
                                     & $120$  & $120$   
                                     & $(119\pm1)^{+5}_{-5}$  \\
\rule[-2mm]{0mm}{0mm}$\Delta_{1600} $&$ \frac{3}{2}^+$ 
                                     & $43$    & $34$ 
                                     & $174$   & $14$    
                                     & $(61\pm26)^{+26}_{-10}$  \\
\rule[-2mm]{0mm}{0mm}$N_{1520} $&$ \frac{3}{2}^-$     
                                     & $131$ & $161$   
                                     & $108$ & $168$    
                                     & $(66\pm6)^{+9}_{-5}$ 
				     & $0$   & $0$   
                                     & $0$   & $0$     &  \\
\rule[-2mm]{0mm}{0mm}$N_{1535} $&$ \frac{1}{2}^-$      
                                     & $336$ & $75$ 
                                     & $462$ & $109$   
                                     & $(67\pm15)^{+55}_{-17}$ 
				     & $64$  & $64$ 
                                     & $64$  & $64$  
                                     & $(64\pm19)^{+76}_{-15}$  \\
\rule[-2mm]{0mm}{0mm}$N_{1650} $&$ \frac{1}{2}^-$      
                                     & $53$   & $5$    
                                     & $87$   & $8$     
                                     & $(109\pm26)^{+36}_{-3}$ 
				     & $113$  & $68$    
                                     & $140$  & $94$
                                     & $(10\pm5)^{+4}_{-1}$ \\
\rule[-2mm]{0mm}{0mm}$N_{1675} $&$ \frac{5}{2}^-$      
                                     & $34$   & $35$    
                                     & $40$   & $52$     
                                     & $(68\pm8)^{+14}_{-4}$ 
				     & $2$  & $4$    
                                     & $3$   & $5$ & \\
\rule[-2mm]{0mm}{0mm}$N_{1700} $&$ \frac{3}{2}^-$      
                                     & $6$   & $6$     
                                     & $7$    & $9$     
                                     & $(10\pm5)^{+3}_{-3}$  
				     & $0$   & $1$     
                                     & $1$   & $1$     & \\
\rule[-2mm]{0mm}{0mm}$\Delta_{1620} $&$ \frac{1}{2}^-$ 
                                     & $26$    & $3$    
                                     & $41$    & $5$     
                                     & $(38\pm8)^{+8}_{-6}$   \\
\rule[-2mm]{0mm}{0mm}$\Delta_{1700} $&$ \frac{3}{2}^-$ 
                                     & $28$    & $29$   
                                     & $20$   & $38$     
                                     & $(45\pm15)^{+20}_{-10}$  \\
\rule[-2mm]{0mm}{0mm}$N_{1680} $&$ \frac{5}{2}^+$     
                                     & $85$  & $85$    
                                     & $149$   & $313$    
                                     & $(85\pm7)^{+6}_{-6}$ 
				     & $0$   & $1$    
                                     & $2$   & $6$        & \\
\rule[-2mm]{0mm}{0mm}$N_{1720} $&$ \frac{3}{2}^+$      
                                     & $377$ & $100$  
                                     & $689$  & $238$ 
                                     & $(23\pm8)^{+9}_{-5}$ 
				     & $15$  & $11$  
                                     & $30$  & $25$      &  \\
\hline
\multicolumn{2}{c}{\rule[-4mm]{0mm}{10mm} $\gamma$} &
 15.365 & 14.635 & 18.015 & 11.868 & & 5.929 & 6.682 & 6.572 & 4.937 & \\
\end{tabular}
\end{table}

Table \ref{tab1} also allows a comparison of the theoretical results to
experimental data as compiled by the Particle Data Group (PDG) \cite{PDG}. 
For the latter there arise
two kinds of uncertainties: First, the total decay width of each resonance is
given by a central value and a lower and upper bound. Second, the partial decay
width has its own uncertainty. In Table \ref{tab1} we quote the value for the
$\pi$ decay widths deduced from the central value of the total width and first
add the uncertainty from the partial decay width itself (numbers inside the
parantheses in the last column). Then we indicate also the range of the total
decay width by an upper and lower bound. We understand that the total uncertainty
in a partial decay width must be estimated by combining both types of
uncertainties (inherent separately in the total and partial widths).

Let us now examine the theoretical results in detail. 
For the $N_{1440}$ $\frac{1}{2}^+$ resonance the SR GBE prediction
is obviously too large, 
whereas the pertinent NR result lies within the experimental error bars. 
The SR OGE result overshoots the
experiment by far, its NR version is also much smaller than the 
SR one and lies just at the lower end of the experimental error bar.
The results for the next $\frac{1}{2}^+$ excitation of the nucleon, 
the $N_{1710}$, show a similar relative
pattern as the ones for the Roper resonance,
though all the values are smaller by about an order of magnitude. 
The fact that for each case, $N_{1440}$ and $N_{1710}$,
the predictions of the SR parametrizations of both the OGE and GBE
models exceed by far their NR counterparts
can be readily understood observing the higher momentum
components present in the SR parametrizations, as compared to the NR ones
(cf. the discussion of the baryon wave functions in the previous 
Subsection). In case of the OGE SR this effect is enhanced by a phase space
that is much too large (due to the bad prediction of the resonance 
energy).

For the $N_{1720}$ $\frac{3}{2}^+$ resonance the results again have similar 
characteristics, with the SR cases drastically
overshooting the experimental data. Here, however, none of the NR 
versions can come close to the rather small experimental width. 
This problem was already encountered in
similar analyses \cite{CAPROB,STAST2} and may hint to a wrong symmetry
assignment (or a strong mixing) of this state.
Only for the $N_{1680}$
$\frac{5}{2}^+$ resonance the GBE CQM produces correct results, both
in its SR and NR versions. In this case the results from both 
variants of the OGE CQM are again too high.

For the negative-parity $N_{1535}$ $\frac{1}{2}^-$ resonance the 
SR results are also much too high, whereas the predictions from the 
NR versions agree with experiment. For the $N_{1650}$ $\frac{1}{2}^-$
the situation is just reversed.
Most remarkably, in all instances the widths of
the $N_{1535}$ resonance are larger than the ones of $N_{1650}$, 
contrary to experiment, where
the $N_{1535}$ width appears to be smaller or is at most as large as the
$N_{1650}$ width (taking into account the experimental uncertainties).
Regarding the $L=1$, $S=\frac{3}{2}$ multiplet $N_{1650}-N_{1675}-N_{1700}$,
one notes that the SR parametrizations give approximately the correct
ratios of these widths, as it is expected from the corresponding
spin-isospin matrix elements. These features are not found for the NR
parametrizations due to the exceedingly small value of the $N_{1650}$
width.

Concerning the negative-parity $N$ excitations, it is interesting to note
that certain resonances are more sensible to the different parametrizations
than others. Specifically, the S-wave resonances $N_{1535}$ and $N_{1650}$
(and likewise also $\Delta_{1620}$) appear to be
'structure-dependent', following the terminology of ref. \cite{KONIUK}.
This behaviour results in
widths sometimes orders of magnitudes apart for different models.
On the other hand, the D-wave resonances $N_{1520}$, $N_{1675}$, and
$N_{1700}$ (and likewise also $\Delta_{1700}$) are found to be
'structure-independent'. Their decay widths are
practically independent of the underlying 
spectroscopic model. These properties can be easily understood in the framework
of the EEM (see ref. \cite{YAOUAN} for a thorough discussion), and evidently
extend to the $^3P_0$ model, which is qualitatively very similar for orbital
excitations.

The decay widths for the $\Delta$ resonances are practically all 
correct for the SR GBE CQM. In case of the other models 
the one or the other shortcoming appears.

\subsection{$\eta$ decays}

Table \ref{tab1} also gives the results for $\eta$ decays. Here we use the same
spatial part for the meson wave function as for $\pi$ decays
but the constant $\gamma$ is adjusted so as to reproduce
the $\eta$ decay width of the $N_{1535}$ resonance. Note that this 
gives values for $\gamma$ about a factor 3 smaller than for the
$\pi$ decays, in contrast to other works \cite{CAPROB}, where the same
value was employed to describe both the $\pi$ and $\eta$ decays.
This has several reasons, the most imminent one being the replacement 
according to
Eq. (\ref{replacement1}). Furthermore, we use an unmixed flavor wave function
for the $\eta$ meson, i.e. a pure flavor octet state. For non-strange decays as
regarded in this work, a possible mixing would only influence the normalization
of this wave function, which can effectively be absorbed into the coupling
constant $\gamma$. Finally, an important contribution comes from our choice of
phase space, as given by Eq. (\ref{width}). We use a fully relativistic
prescription and experimental values for the meson masses, 
in contrast to ref. \cite{CAPROB}, where a much higher,
''effective'' value for the pion mass was employed. A quick estimate of the
magnitudes of these three effects shows indeed that we end up with
about a factor of 3 difference in the constant $\gamma$ between
$\pi$ and $\eta$ decays.

The $\eta$ widths of the Roper resonance $N_{1440}$ for the GBE 
parametrizations (NR as well as SR) are rigourously zero,
since in both cases the theoretically predicted masses lie below the $\eta$
threshold, in accordance with experiment. 
For the OGE parametrizations, the decay $N_{1440}\rightarrow N\eta$ is
possible, the corresponding widths remain rather small, however.

In total, there are four resonances predicted with considerable branching
ratios in the $\eta$ decay channel. Only for the $N_{1535}$ and 
$N_{1650}$ resonances one can compare to experiment, since these are the
only ones with an experimental width assigned by the PDG \cite{PDG}.
The relative magnitudes of the experimental decay widths in both of 
these cases are missed by all theoretical models. This is again 
reminiscent of the EEM, where a similar effect is found. One may expect
that the decays of these resonances are quite sensitive to 
spin-orbit and/or tensor forces in the quark-quark interaction. The
inclusion of these force components would probably improve the description
of both $N\pi$ and $N\eta$ decays for these resonances.

In addition to $N_{1535}$ and $N_{1650}$, also the widths of the
$N_{1710}$ and the $N_{1720}$ resonances come out appreciably large. 
The PDG does not quote any experimental data for these states. This
does not necessarily mean that their widths are vanishing or too
small to be measured. It may simply be the case that
experimental ambiguities do not (yet) allow for a reliable 
determination. In fact, there are single partial-wave analyses
that assign an appreciable $\eta$ decay width, for example, also
to the $N_{1710}$, see ref. \cite{BATVRA}.

\subsection{Influences of the meson wave function}

The modified $^3P_0$ decay model has two decisive ingredients: 
the pair-creation strength $\gamma$ and the parameter determining
the extension of the meson wave function.
While the former is merely a multiplicative constant, which may be
suitably chosen to scale the overall strength of all decays, the latter
is a nonlinear parameter, which may also alter the qualitative features
of various predictions. In the following we consider certain different
choices of the meson wave functions and examine their influences
on the decay widths.

In Table \ref{tab3} we show results for decay widths when employing
a Yukawa-like meson wave function, as given by Eq. (\ref{yukawa}),
producing the same meson size as the Gaussian parametrization before.
We have adjusted the parameter $\gamma$ again to fit the
$\Delta$ and $N_{1535}$ widths for $N\pi$ and $\eta$ decays, 
respectively. However, as compared to Table \ref{tab1},
the values change only little in this case.

\begin{table}[ht]
\caption{\label{tab3}
Same as Table \protect\ref{tab1} but using a Yukawa-type meson wave
function with 
$r_{\pi}=r_{\eta}=0.565$ fm.}
\begin{tabular}{cccccccccccc}
\rule[-4mm]{0mm}{10mm} $N^*$ & $J^{\pi}$& 
\multicolumn{5}{c}{$\Gamma(N^*\rightarrow N\pi$) [MeV]} &
\multicolumn{5}{c}{$\Gamma(N^*\rightarrow N\eta$) [MeV]} \\
\hline
\rule[-4mm]{0mm}{10mm} &&  GBE SR & GBE NR & OGE SR & OGE NR & exp.
&  GBE SR & GBE NR & OGE SR & OGE NR & exp.\\
\hline
\rule[-2mm]{0mm}{7mm}$N_{1440} $&$ \frac{1}{2}^+$     
                                  & $528$  & $363$  
                                  & $1015$  & $204$
                                  & $(227\pm18)^{+70}_{-59}$
				  &&& $6$ & $15$ &\\
\rule[-2mm]{0mm}{0mm}$N_{1710} $&$ \frac{1}{2}^+$      
                                  & $59$    & $10$    
                                  & $179$    & $7$
				  & $(15\pm5)^{+30}_{-5}$
				  & $32$ & $7$ & $51$ & $14$ &\\
\rule[-2mm]{0mm}{0mm}$\Delta_{1232} $&$ \frac{3}{2}^+$ 
                                  & $120$  & $120$  
                                  & $120$  & $120$
				  & $(119\pm1)^{+5}_{-5}$ 
				  &&&&&\\
\rule[-2mm]{0mm}{0mm}$\Delta_{1600} $&$ \frac{3}{2}^+$ 
                                  & $41$    & $49$    
                                  & $142$    & $11$
				  & $(61\pm26)^{+26}_{-10}$
				  &&&&&\\
\rule[-2mm]{0mm}{0mm}$N_{1520} $&$ \frac{3}{2}^-$     
                                  & $140$ & $225$ 
                                  & $109$ & $187$
				  & $(66\pm6)^{+9}_{-5}$
				  & $0$ & $1$ & $0$ & $1$ &\\
\rule[-2mm]{0mm}{0mm}$N_{1535} $&$ \frac{1}{2}^-$      
                                  & $251$   & $31$   
                                  & $412$   & $61$
				  & $(67\pm15)^{+55}_{-17}$
				  & $64$  & $64$ 
                                  & $64$  & $64$  
                                  & $(64\pm19)^{+76}_{-15}$\\
\rule[-2mm]{0mm}{0mm}$N_{1650} $&$ \frac{1}{2}^-$      
                                  & $39$    & $1$    
                                  & $78$    & $3$
				  & $(109\pm26)^{+36}_{-3}$
				  & $110$ & $59$ & $138$ & $86$ 
				  &$(10\pm5)^{+4}_{-1}$\\
\rule[-2mm]{0mm}{0mm}$N_{1675} $&$ \frac{5}{2}^-$      
                                  & $35$   & $42$   
                                  & $40$   & $55$
				  & $(68\pm8)^{+14}_{-4}$
				  & $3$ & $6$ & $3$ & $7$ &\\
\rule[-2mm]{0mm}{0mm}$N_{1700} $&$ \frac{3}{2}^-$      
                                  & $6$   & $7$   
                                  & $7$   & $9$
				  & $(10\pm5)^{+3}_{-3}$ 
				   & $1$ & $1$ & $1$ & $1$ &\\
\rule[-2mm]{0mm}{0mm}$\Delta_{1620} $&$ \frac{1}{2}^-$ 
                                  & $20$    & $1$    
                                  & $39$    & $2$
				  & $(38\pm8)^{+8}_{-6}$
				  &&&&&\\
\rule[-2mm]{0mm}{0mm}$\Delta_{1700} $&$ \frac{3}{2}^-$ 
                                  & $28$    & $35$    
                                  & $21$    & $40$
				  & $(45\pm15)^{+20}_{-10}$
				  &&&&&\\
\rule[-2mm]{0mm}{0mm}$N_{1680} $&$ \frac{5}{2}^+$     
                                  & $98$  & $144$  
                                  & $158$ & $379$
				  & $(85\pm7)^{+6}_{-6}$
				  & $1$ & $2$ & $2$ & $10$ &\\
\rule[-2mm]{0mm}{0mm}$N_{1720} $&$ \frac{3}{2}^+$      
                                  & $276$ & $58$  
                                  & $545$  & $132$
				  & $(23\pm8)^{+9}_{-5}$
				  & $14$ & $11$ & $25$ & $22$ &\\
\hline
\multicolumn{2}{c}{\rule[-4mm]{0mm}{10mm} $\gamma$} &
15.931 & 14.741 & 18.854 & 11.961 & & 6.608 & 7.022 & 7.056 & 5.469 &\\
\end{tabular}
\end{table}

By comparing the results in Tables \ref{tab1} and \ref{tab3} it is immediately
seen that the specific form of the meson wave function has
only a minor influence on the predictions of the decay widths for the
$\pi$ as well as $\eta$ decay modes. The qualitative
features remain essentially unchanged. We have also performed calculations
with the exact meson wave function produced by the potential of
Bhaduri et al. (as shown in Fig.\ \ref{ygbwf}). They confirm the conclusion
that the type of meson wave function is not decisive, provided its 
extension (meson radius) is kept the same.

We now focus the attention on the dependence of the results on the
size of the meson. The meson wave functions employed in Tables \ref{tab1}
and \ref{tab3} both correspond to a radius of $r_{\pi}=0.565$ fm.
In the limit $r_{\pi} \rightarrow 0$ one
expects to reproduce the results of the EEM. 
Thus it is interesting to look at an intermediate regime.
Table \ref{tab4} gives the decay widths for the same case as
in Table \ref{tab1}, but for a Gaussian-type
wave function leading to a meson radius as small as 0.36 fm.

\begin{table}[ht]
\caption{\label{tab4}
Same as Table \protect\ref{tab1} but using a Gaussian-type meson wave
function with $r_{\pi}=r_{\eta}=0.36$ fm. }

\begin{tabular}{cccccccccccc}
\rule[-4mm]{0mm}{10mm}$N^*$ & $J^{\pi}$ & 
\multicolumn{5}{c}{$\Gamma(N^*\rightarrow N\pi$) [MeV]} & 
\multicolumn{5}{c}{$\Gamma(N^*\rightarrow N\eta$) [MeV]} \\
\hline
\rule[-4mm]{0mm}{10mm}&&  GBE SR & GBE NR & OGE SR & OGE NR & exp.
&  GBE SR & GBE NR & OGE SR & OGE NR & exp. \\
\hline\hline
\rule[-2mm]{0mm}{0mm}$N_{1440} $&$ \frac{1}{2}^+$     
                                     & $240$    & $69$  & 
                                       $546$   & $44$ 
                                     & $(227\pm18)^{+70}_{-59}$ 
				     &  &  & 2 & 4 &\\
\rule[-2mm]{0mm}{0mm}$N_{1710} $&$ \frac{1}{2}^+$      
                                     & $6$    & $13$     &
                                       $63$   & $26$     
                                     & $(15\pm5)^{+30}_{-5}$ 
				     & $9$  & $1$  
                                     & $18$   & $4$ &  \\
\rule[-2mm]{0mm}{0mm}$\Delta_{1232} $&$ \frac{3}{2}^+$ 
                                     & $120$   & $120$    & 
                                       $120$  & $120$   
                                     & $(119\pm1)^{+5}_{-5}$
				     &&&&&\\  
\rule[-2mm]{0mm}{0mm}$\Delta_{1600} $&$ \frac{3}{2}^+$ 
                                     & $0$     & $2$     &
                                       $24$    & $63$     
                                     & $(61\pm26)^{+26}_{-10}$
				     &&&&&\\  
\rule[-2mm]{0mm}{0mm}$N_{1520} $&$ \frac{3}{2}^-$     
                                     & $89$   & $88$     &
                                       $81$  & $137$    
                                     & $(66\pm6)^{+9}_{-5}$  
				     & $0$   & $0$   
                                     & $3$   & $0$     & \\
\rule[-2mm]{0mm}{0mm}$N_{1535} $&$ \frac{1}{2}^-$      
                                     & $584$   & $106$  &
                                       $953$  & $195$  
                                     & $(67\pm15)^{+55}_{-17}$ 
				     & $64$  & $64$ 
                                     & $64$  & $64$  
                                     & $(64\pm19)^{+76}_{-15}$ \\
\rule[-2mm]{0mm}{0mm}$N_{1650} $&$ \frac{1}{2}^-$      
                                     & $122$    & $14$    &
                                       $227$   & $28$    
                                     & $(109\pm26)^{+36}_{-3}$ 
				     & $128$  & $80$    
                                     & $156$  & $109$
                                     & $(10\pm5)^{+4}_{-1}$ \\
\rule[-2mm]{0mm}{0mm}$N_{1675} $&$ \frac{5}{2}^-$      
                                     & $26$    & $22$     &
                                       $32$   & $46$     
                                     & $(68\pm8)^{+14}_{-4}$  
				     & $1$  & $2$    
                                     & $1$  & $3$     & \\
\rule[-2mm]{0mm}{0mm}$N_{1700} $&$ \frac{3}{2}^-$      
                                     & $4$     & $4$      &
                                       $5$    & $8$     
                                     & $(10\pm5)^{+3}_{-3}$ 
				     & $0$   & $0$     
                                     & $0$   & $1$     &  \\
\rule[-2mm]{0mm}{0mm}$\Delta_{1620} $&$ \frac{1}{2}^-$ 
                                     & $61$    & $8$    &
                                       $106$   & $16$    
                                     & $(38\pm8)^{+8}_{-6}$  
				     &&&&&\\
\rule[-2mm]{0mm}{0mm}$\Delta_{1700} $&$ \frac{3}{2}^-$ 
                                     & $21$     & $18$    &
                                       $17$   & $34$     
                                     & $(45\pm15)^{+20}_{-10}$  
				     &&&&&\\
\rule[-2mm]{0mm}{0mm}$N_{1680} $&$ \frac{5}{2}^+$     
                                     & $50$    & $41$     &
                                       $93$   & $226$    
                                     & $(85\pm7)^{+6}_{-6}$ 
				     & $0$   & $0$    
                                     & $1$   & $3$        & \\
\rule[-2mm]{0mm}{0mm}$N_{1720} $&$ \frac{3}{2}^+$      
                                     & $489$ & $85$ &
                                       $1063$  & $352$ 
                                     & $(23\pm8)^{+9}_{-5}$  
				     & $12$  & $8$  
                                     & $24$  & $23$      & \\
\hline
\multicolumn{2}{c}{\rule[-4mm]{0mm}{10mm} $\gamma$} &
20.575 & 20.695 & 22.699 & 17.997 &  & 6.844 & 10.060 & 6.430 & 6.619 &\\
\end{tabular}
\end{table}

First we note that the values for the constant $\gamma$ obtained in this case
are considerably larger than before. 
This is understandable, since in order to recover the
results of the pointlike meson limit, one has to compensate for the 
effect of the $\delta$ function, which then replaces the meson wave function.
In particular, for the Gaussian form of Eq. (\ref{gauss}) one has the relation

\begin{equation}
\label{pointlike}
(2 \pi)^{\frac{3}{2}} \delta(\vec{r}) = \lim_{R \rightarrow 0} 
\left( \frac{\pi}{R^2} \right)^{\frac{3}{4}} \Psi_G (\vec{r}).
\end{equation}

Most of the results for the decay widths are now rather different from before.
They follow the general trend towards the predictions typical for the EEM.
One of the characteristic results of the EEM is the extremely small decay
width of the Roper resonance, as the first radial excitation of the 
nucleon; it is due to the orthogonality of the initial and final-state
wave functions, which is strikingly felt in case of the EEM.
The results of Table \ref{tab4} show the corresponding trend rather clearly:
for all spectroscopic models the $N_{1440}$ widths come out at least a
factor of 2 smaller than before, while one is still rather far away from
the pointlike limit.

Concerning the $\eta$ decays one observes that the differences in the
widths between the $N_{1535}$ and $N_{1650}$ resonances now increase
in all cases. Again this follows the (unpleasant) trend towards
the predictions typical for the EEM. As a result it appears favourable
to use a decay model that permits the use of meson wave functions with
finite extensions.

\section{Summary and Conclusion}

In this work we investigated the theoretical description of $\pi$ and $\eta$ 
decays for $N$ and $\Delta$ resonances. In the first instance we were 
interested in the predictions of the specific
chiral constituent-quark model whose hyperfine interaction is based on 
GBE dynamics \cite{GLOZ1,FRASC}. A detailed comparison to the modern
experimental data base \cite{PDG} is provided. 
We also studied the results relative to the predictions
by a traditional CQM \cite{BHAD} based on OGE but relying on the same
type of force components as the GBE CQM. Furthermore we investigated 
the differences between a semirelativistic and a nonrelativistic 
description of the baryon states for both types of CQMs.
For the decay mechanism a modified version of the $^3P_0$ model
\cite{CANO2} was employed. We also examined the sensitivity of the results
on the ingredients entering the decay operator, notably the analytical form
and the extension of the meson wave functions.

From the present results it is still difficult to draw definite conclusions
about the quality of the  wave functions stemming from different CQMs.
In fact, the various decay widths seem to be more determined by the choice
of the SR or NR parametrizations rather than by the use of either 
type of dynamics, GBE or OGE. At this stage, we find a number of gross
qualitative features that have been observed already before in similar
studies along the classical $^3P_0$ decay model.

It should be recalled that here we have not 
included spin-orbit or tensor forces into the quark-model 
Hamiltonians, especially because these force components are not yet 
provided by the published versions of the GBE CQM and we wanted to 
produce a consistent comparison with the other type of dynamics, 
namely the one resulting from OGE. Some decay widths are certainly
sensitive to tensor and spin-orbit components in the wave functions.
In this respect it may have been somewhat premature to make a 
comparison with experimental data at this stage.

In any case, our study reveals (and confirms previous such findings)
that the description of strong decays of baryon resonances
within present CQMs is not yet fully satisfactory. The reasons for the
persisting difficulties may on the one hand
reside in the baryon wave functions, which are probably not yet realistic
enough. On the other hand one must realize that the $^3P_0$ decay model
may also fall short as it is based on intuitive grounds and lacks a
firm theoretical foundation. A consistent microscopic description of
the strong-decay processes within the framework of CQMs thus remains a 
challenging task. One can think of a number of improvements to be 
done. For example, the proper inclusion of relativistic effects 
appears mandatory. The ultimate goal would, of course, be a
unified description of the resonance spectra and the hadronic,
as well as electromagnetic, transitions
with the same dynamical scheme.

\bigskip
{\bf Acknowledgements}
The authors are indebted to Fl. Stancu, D. Rebreyend, and J.P. Bocquet 
for useful discussions.
This work was supported by the Scientific-Technical Agreement 'Amad\'ee' between
Austria and France under contract number II.9 and by the TMR contract ERB
FMRX-CT96-0008.



\begin{thebibliography}{99}

\bibitem{FBSSUP} See, e.g., {\it $N^*$ Physics and Nonperturbative Quantum
                 Chromodynamics}
                 (Proceedings of the Joint ECT*/JLAB Workshop, Trento, 1998),
                 edited by S. Simula B. Saghai, N.C. Mukhopadhyay, 
                 and V.D. Burkert, Few-Body Syst. Suppl. {\bf 11} (1999).
\bibitem{AOK99}  S. Aoki et al., Phys. Rev. Lett. {\bf 82}, 4392 (1999).
\bibitem{SKUWIL} J.I. Skullerud and A.G. Williams, hep-lat/0007028.
\bibitem{GLOZ1}  L.Ya. Glozman, W. Plessas, K. Varga, and R.F. 
                 Wagenbrunn, Phys. Rev. D {\bf 58}, 094030 (1998).
\bibitem{GLOZ2}  L.Ya. Glozman and D.O. Riska, Phys. Rep. {\bf 268}, 263
                 (1996).
\bibitem{GLOZ3}  L.Ya. Glozman, Z. Papp, W. Plessas, K. Varga, and 
                 R.F. Wagenbrunn, Phys. Rev. C {\bf 57}, 3406 (1998).
\bibitem{SUZUK}  Y. Suzuki and K. Varga, {\it Stochastic Variational 
                 Approach to Quantum-Mechanical Few-Body Problems},
		         (Springer, Berlin, 1998).
\bibitem{CANO2}  F. Cano, P. Gonz\'alez, S. Noguera, and B. Desplanques,
                 Nucl. Phys. {\bf A603}, 257 (1996).
\bibitem{BHAD}   R.K. Bhaduri, L.E. Cohler, and Y. Nogami, 
                 Nuovo Cim. {\bf 65A}, 376 (1981).                
\bibitem{CARL}   J. Carlson, J. Kogut, and V.R. Pandharipande, 
                 Phys. Rev. D {\bf 27}, 233 (1983); ibid. {\bf 28}, 2807
                 (1983).
\bibitem {FRASC} L.Ya. Glozman, Z. Papp, W. Plessas, K. Varga, and 
                 R.F. Wagenbrunn, Nucl. Phys. {\bf A623}, 90c (1997).
\bibitem{PDG}    D.E. Groom et al. (Particle Data Group), 
                 Eur. Phys. Jour. {\bf C15}, 1 (2000). 
\bibitem{BECCHI} C. Becchi and G. Morpurgo, Phys. Rev. {\bf 149}, 1284 (1966).
\bibitem{MITRA}  A. Mitra and M. Ross, Phys. Rev. {\bf 158}, 1630 (1967).
\bibitem{FAIMAN} D. Faiman and A.W. Hendry, Phys. Rev. {\bf 173}, 1720 (1968);
                 Phys. Rev. Lett. {\bf 44}, 845 (1980).
\bibitem{STAST1} Fl. Stancu and P. Stassart, Phys. Rev. C {\bf 39}, 343 (1996).
\bibitem{GLOZ4}  L. Ya. Glozman W. Plessas, L. Theussl, K. Varga, and 
                 R.F. Wagenbrunn, $\pi$N Newsletter {\bf 14}, 99 (1998).
\bibitem{YAOUAN} A. Le Yaouanc, Ll. Oliver, O. P\`ene, and J.-C. Raynal,
                 {\it Hadron Transitions in the Quark Model},
                 (Gordon and Breach Science Publishers, New York, 1988).
\bibitem{PKPLE}  Z. Papp, A. Krassnigg, and W. Plessas, Phys. Rev. C
                 {\bf 62}, 044004 (2000). 
\bibitem{CAPROB} S. Capstick and W. Roberts, Phys. Rev. D {\bf 47}, 1994 (1993).
\bibitem{STAST2} Fl. Stancu and P. Stassart, Phys. Rev. D {\bf 38}, 233 (1988).
\bibitem{KONIUK} R. Koniuk and N. Isgur, Phys. Rev. D {\bf 21}, 1868 (1980).
\bibitem{BATVRA} M. Batinic, I. Slaus, A. Svarc, and B.M.K. Nefkens, 
                 Phys. Rev. C {\bf 51}, 2310 (1995);
                 T.P. Vrana, S.A. Dytman, and T.-S.H. Lee, 
		         Phys. Rep. {\bf 328}, 181 (2000).
\end{thebibliography}
\end{document}